\documentclass[reprint,amsmath,amssymb,aps,nofootinbib]{revtex4-1}
\usepackage{dcolumn}
\usepackage{bm}
\usepackage{amsmath,amsthm,amsfonts,float,latexsym}
\usepackage[pdftex]{hyperref}
\begin{document}
\title{Electrovac universes with a cosmological constant}

\author{Camilo Posada}
 \affiliation{%
Department of Physics and Astronomy\\
University of South Carolina\\
Columbia, SC 29208 USA.}
\author{Davide Batic}%
\affiliation{%
 Department of Mathematics\\
 The University of The West Indies\\
 Mona, Jamaica\\
}%
\date{\today}%
\begin{abstract}
We present the extension of the Einstein-Maxwell system called electrovac universes by 
introducing a cosmological constant $\Lambda$. 
In the absence of the $\Lambda$ term, the crucial equation in 
solving the Einstein-Maxwell system is the Laplace equation. 
The cosmological constant modifies this equation to become in a nonlinear partial differential equation 
which takes the form $\Delta U=2\Lambda U^3$. We offer special solutions of this equation.
\end{abstract}
\maketitle
\section{Introduction}
\vskip-0.05in
The Einstein-Maxwell system has always been an area of active interest due to its physical relevance and the mathematical challenge which it presents. Not only has it unique features \cite{rainich}, but also offers a class of interesting solutions. One of them is known under the name of electrovac universes \cite{majum, pap, synge, raychaud} whose sub-class, the conformastat metric\cite{majum, pap, synge, raychaud}, leads to elegant solutions in which the metric elements are given by solutions of Laplace equation as long as the cosmological constant is zero. Under the same conditions, the static electric potential can be obtained by solving a polynomial equation of second order. The recent interest in the cosmological constant as an explanation of the accelerated universe \cite{accel} makes the introduction of the cosmological constant also in the Einstein-Maxwell system relevant. Indeed, the cosmological constant affects not only the cosmological aspect of Einstein's General Relativity, but also local phenomena \cite{lambda}. It seems therefore timely to investigate the role of $\Lambda$ in Einstein-Maxwell system. We will first show that the introduction of $\Lambda$ changes the mathematical aspect of the electrovac universes: the linear Laplace equation and the algebraic one for the electric potential become both non-linear partial differential equations. In deriving these results one needs some knowledge and results already derived in literature. In section two and three we will review these issues which in section four will be used to arrive at the main result. In section four we dwell on some particular solutions of the generalized Laplace equation with $\Lambda$. 
\section{Static universes in conformastat form}
Per definition, a general static universe is represented by the following metric
\begin{equation}\label{conformastat}
ds^2=-V^2(x^i)dt^2+U^2(x^i)dx^idx^i
\end{equation} 
which Synge\cite{synge} calls \textit{conformastat}. Note that the metric elements in (\ref{conformastat}) depends only on spatial coordinates. Therefore the metric (\ref{conformastat}) is invariant under $x^0\to x^0+const$. We can find a Killing vector $\xi^{\alpha}$ associated with this symmetry, namely
\begin{equation}
{\bf\xi\cdot u}=\text{g}_{tt}\xi^{\alpha}u^{t}=-V^2u^{t}=const.
\end{equation}
\noindent where $\xi^{\alpha}=(1,0,0,0)$ in the basis $(t,x,y,z)$. The Christoffel symbols can be readily calculated to be\footnote{The Christoffel symbols, the components of the Ricci tensor and the Ricci scalar, were calculated using the Mathematica Notebook available in reference \cite{hartle}}
\begin{eqnarray}
\Gamma_{i0}^{0}=V^{-1}V_{,_{i}}\,\,\,\,;\,\,\,\,\Gamma_{00}^{i}=U^{-2}VV_{,_{i}}\nonumber\\\Gamma_{ij}^{i}=U^{-1}U_{,_{j}}\,\,\,\,;\,\,\,\,\Gamma_{jj}^{i}=-U^{-1}U_{,_{i}}\,\,\text{for}\,\,i\neq j
\end{eqnarray}
\noindent The spatial part of the Ricci tensor is
\begin{eqnarray} \label{spatial}
R_{ij}=U^{-1}(U_{,_{ij}}+\delta_{ij}U_{,_{kk}})-2U^{-2}U_{,_{i}}U_{,_{j}}+V^{-1}V_{,_{ij}}\nonumber\\-(UV)^{-1}(U_{,_{i}}V_{,_{j}}+U_{,_{j}}V_{,_{i}})+(UV)^{-1}\delta_{ij}U_{,_{k}}V_{,_{k}}
\end{eqnarray}
\noindent whereas the temporal part takes the form
\begin{equation} \label{temporal}
R_{00}=-VU^{-2}(V_{,_{kk}}+U^{-1}U_{,_{k}}V_{,_{k}})
\end{equation}
\noindent This allows us to calculate the Ricci scalar $R=g^{\mu\nu}R_{\mu\nu}$ as
\begin{eqnarray} \label{scalar}
R=4U^{-3}\left(U_{,_{kk}}-\frac{1}{2}U^{-1}U_{,_{k}}U_{,_{k}}\right)\nonumber\\+2U^{-2}V^{-1}(V_{,_{kk}}+U^{-1}U_{,_{k}}V_{,_{k}})
\end{eqnarray}
\noindent These results are valid for any metric of the \textit{conformastat} type. Before turning our attention to the Einstein-Maxwell system, it is of some interest to see what results we obtain if we specialize to the vacuum case i.e.
\begin{equation} \label{vacuum}
R_{\mu\nu}=0
\end{equation}
\noindent Equation (\ref{vacuum}) implies 
\begin{equation} \label{vacuum2}
R=0
\end{equation} 
\noindent and also, with the help of (\ref{temporal}), 
\begin{equation} \label{Poisson}
V_{,_{kk}}+U^{-1}U_{,_{k}}U_{,_{k}}=0
\end{equation}
Therefore equation (\ref{scalar}) tells us that
\begin{equation}
U_{,_{kk}}-\frac{1}{2}U^{-1}U_{,_{k}}U_{,_{k}}=0
\end{equation}
which can be cast into a Laplace equation
\begin{equation} \label{Laplace1}
(\sqrt{U})_{,_{kk}}=0
\end{equation}
\noindent Hence all $\sqrt{U}$ must be \textit{harmonic functions}. Thus, knowing $U$ we can obtain $V$ by solving the Poisson equation (\ref{Poisson}). One well known solution is, of course, the Schwarzschild solution. In isotropic coordinates the latter is given by
\begin{equation} \label{isotropic}
ds^2=-\left(\frac{1-\xi}{1+\xi}\right)^2dt^2+(1+\xi)^4(d\rho^2+\rho^2d\Omega^2)
\end{equation}
\noindent where $\xi=\frac{m}{2\rho}$. Comparing (\ref{isotropic}) with (\ref{conformastat}) we conclude
that
\begin{equation}
U=(1+\xi)^2\,\,\,\,\,;\,\,\,\,\,V=\frac{1-\xi}{1+\xi}
\end{equation}
\noindent with $\rho^2=x^ix^i$. It is an easy exercise to show now that $\sqrt{U}$ is a harmonic function in the isotropic coordinates.\\
\newline
So far, we have specialized on the vacuum case. We can go further in our development, by introducing the condition $UV=1$ such that the metric reads now
\begin{equation}
ds^2=-U^{-2}dt^2+U^{2}dx^{i}dx^{j}\delta_{ij}
\end{equation}
\noindent Equation (\ref{spatial}) can be written as
\begin{align*}
R_{ij} &= U^{-1}(U_{,_{ij}}+\delta_{ij}U_{,_{kk}})-2U^{-2}U_{,_{i}}U_{,_{j}}+U(U^{-1})_{,_{ij}}\\
& -U_{i}(U^{-1})_{,_{j}}-U_{j}(U^{-1})_{,_{i}}+\delta_{ij}U_{,_{k}}(U^{-1})_{,_{k}}
\end{align*}
\noindent which simplifies considerably to
\begin{equation}
R_{ij}=\delta_{ij}U^{-1}(U_{,_{kk}}-U^{-1}U_{,k}U_{,k})+2U^{-2}U_{,_{i}}U_{,_{j}}
\end{equation} 
by using $(U^{-1})_{,_{ij}}=2U^{-3}U_{,_{j}}U_{,_{i}}-U^{-2}U_{,_{ij}}$.We give also the corresponding spatial components of the Ricci tensor and the Ricci scalar:
\begin{equation}
R_{00}=U^{-5}(U_{,_{kk}}-U^{-1}U_{,k}U_{,k})
\end{equation}
\begin{equation} \label{important}
R=2U^{-3}U_{,_{kk}}
\end{equation}
\noindent Note that in the last equations we have not used the Einstein equation. As a consequence, equation (\ref{important}) represents a general result for the function $U$ if $R$ is known (e.g. vacuum case, conformal energy-momentum tensor with $T^{\mu}_{\mu}=0$ etc.). 
\section{Electrovac universe}
The scenario is as follows. We have a static electric charge located somewhere in the spacetime under consideration. This implies that in the exterior region
we have only an electric field, but no matter. This is called by Synge the \textit{electrovac universe} \cite{synge}. Such a situation can be understood as a generalization of the Reissner-Nordstr\"om metric who consider the spherically symmetric case. Let us first choose the metric to be of the form
\begin{equation} \label{electrovac}
ds^2=-V^2(x^i)dt^2+h_{ij}(x^k)dx^idx^j
\end{equation}
We will first derive results in realm of this metric before specializing, at the end, to the conformastat case (\ref{conformastat}). We are looking for the solutions of the Einstein equation (here still with $\Lambda=0$)
\begin{equation}
R_{\mu\nu}-\frac{1}{2}R\text{g}_{\mu\nu}=-8\pi T_{\mu\nu}
\end{equation}
\noindent where the source $T_{\mu \nu}$ is the electromagnetic energy-momentum tensor
\begin{equation} \label{energy}
T_{\mu\nu}=\frac{1}{4}\text{g}_{\mu\nu}F_{\alpha\beta}F^{\alpha\beta}-F_{\mu\alpha}F_{\nu}^{\alpha}
\end{equation}
and the Maxwell equations in vacuum
\begin{equation} \label{Maxwell}
F^{\mu\nu}_{;_{\nu}}=F^{\mu\nu}_{,_{\nu}}+\Gamma_{\alpha\nu}^{\nu}F^{\mu\alpha}+\Gamma_{\alpha\nu}^{\mu}F^{\alpha\nu}=0
\end{equation}
under the condition that the system is purely electrostatic. This implies that there is only one component of 
\begin{equation}
F_{\mu\nu}=A_{\nu,_{\mu}}-A_{\mu,_{\nu}}
\end{equation}
\noindent which is non-zero\footnote{The mathematically possible case $F_{23}=-F_{32}\neq 0$ would indicate a magnetic monopole which we leave out of our discussion.}:
\begin{equation} \label{Maxwell2}
F_{0i}=-A_{0,_{i}}=-\phi_{,_{i}}
\end{equation}
\noindent where $\phi$ is the electric potential. In terms of this potential the spatial components of the electromagnetic tensor can be obtained as
\begin{equation} \label{energy2}
T_{ij}=V^{-2}\left(\frac{1}{2}h_{ij}\Delta_{1}\phi-\phi_{,_{i}}\phi_{,_{j}}\right)
\end{equation}
where we have defined
\begin{equation} \label{def1}
\Delta_{1}\phi\equiv h^{ij}\phi_{,i}\phi_{,j}
\end{equation}
The temporal components are simply
\begin{equation} \label{energy3}
T_{00}=-\frac{1}{2}h^{ij}\phi_{,i}\phi_{,j}=-\frac{1}{2}\Delta_{1}\phi
\end{equation}
The explicit form of the energy-momentum tensor can be now used to write down the Einstein equation as
\begin{equation} \label{Einstein2}
R_{ij}=-8\pi V^{-2}\left(\frac{1}{2}h_{ij}\Delta_{1}\phi-\phi_{,i}\phi_{,j}\right)
\end{equation}

\begin{equation} \label{Einstein3}
R_{00}=4\pi\Delta_{1}\phi = V\Delta_{2}V
\end{equation}
In the above we used the traceless condition of the electromagnetic tensor such that the Einstein equation takes the form $R_{\mu\nu}=-8\pi T_{\mu\nu}$. Here we introduced a new definition, namely \begin{equation} \label{def2}
\Delta_{2}\phi\equiv h^{ij}\phi_{\vert\vert_{ij}}
\end{equation} 
\noindent The double vertical lines indicates the covariant derivative with respect to the spatial metric $h_{ij}$. Note that the remaining components of the Einstein equation, $R_{i0}=-8\pi T_{i0}$ are identically satisfied.\\ 

\noindent It is clear that the only relevant component of the Maxwell equation is
\begin{equation} \label{zero}
F^{0i}_{;_{i}}=F^{0i}_{,_{i}}+\Gamma^{i}_{ik}F^{0k}+\Gamma_{0i}^{0}F^{0i}=0
\end{equation}
\noindent with
\begin{equation}
F^{0i}=\text{g}^{00}h^{ij}F_{0j}=V^{-2}h^{ij}\phi_{,_{j}}
\end{equation}
\noindent one obtains easily 
\begin{equation} \label{one}
F^{0i}_{,_{i}}=V^{-2}(-2V^{-1}V_{,_{i}}h^{ij}\phi_{,_{j}}+\phi_{,_{j}}h^{ij}_{,_{i}}+h^{ij}\phi_{,_{ij}})
\end{equation}
On the other hand we have
\begin{equation}\label{two}
\Gamma_{ik}^{i}F^{0k}=\frac{1}{2}V^{-2}h^{lm}(h_{lm_{,k}})(h^{kj}\phi_{,_{j}})
\end{equation}

\begin{equation}
\Gamma_{0i}^{0}F^{0i}=V^{-3}h^{ij}V_{,_{i}}\phi_{,_{j}}
\end{equation}
\noindent Using (\ref{one}) and (\ref{two}) equation (\ref{zero}) takes the form
\begin{eqnarray}\label{three}
F^{0i}_{;_{i}} &= V(h^{ij}\phi_{,_{ij}}+h^{ij}_{,_{i}}\phi_{,_{j}})+\frac{1}{2}V\left(h^{lm}(h_{lm})_{,_{k}}h^{kj}\phi_{,_{j}}\right)\nonumber\\
& -h^{ij}V_{,_{i}}\phi_{,_{j}}=0
\end{eqnarray}
\noindent which can be simplified further noticing that 
$h^{ij}_{,_{i}}=(h_{ij})^{-1}_{,_{i}}=-h_{ij}^{-2}h_{ij_{,_{i}}}$
and 
\begin{equation}
\phi_{\vert\vert_{ij}}=(\phi_{,_{i}})_{;_{j}}=\phi_{,_{ij}}-\Gamma_{ij}^{\gamma}\phi_{\gamma}=
\phi_{,_{ij}}-\frac{1}{2}h^{km}h_{km_{,_{i}}}\phi_{,_{j}}\nonumber
\end{equation}
\noindent Making use of the definition (\ref{def2}) we can rewrite equation (\ref{three}) in an elegant form, namely
\begin{equation} \label{Maxwell3}
F^{0i}_{;_{i}}=V\Delta_{2}\phi-h^{ij}V_{,_{i}}\phi_{,_{j}}=0
\end{equation} 
\noindent In particular, we are interested in solutions where $V$ and $\phi$ are functionally related $V=V(\phi)$\cite{majum}. This condition, allows us to write the following
\begin{eqnarray} \label{aid}
V_{\vert\vert_{ij}}=V'\phi_{\vert\vert_{ij}}+V''\phi_{,_{i}}\phi_{,_{i}}\,\,\,\,\, ; \,\,\,\,\,V'=\frac{dV}{d\phi}\,\,\,\,\,\nonumber\\V''=\frac{d^2V}{d\phi^2}\,\,\,\,\, ; \,\,\,\,\,V_{,_{i}}=V'\phi_{,_{i}}
\end{eqnarray} 
\noindent To summarize, we are looking for $V$, $\phi$ y $h_{ij}$ such that the Einstein equation (\ref{Einstein2}), (\ref{Einstein3}) and the Maxwell equation (\ref{Maxwell3}) are satisfied. Concentrating first on (\ref{Einstein3}) and (\ref{Maxwell3}) this means that we have to solve
\begin{equation} \label{Einstein3.1}
V\Delta_{2}V-4\pi\Delta_{1}\phi=0 
\end{equation}
and
\begin{equation} \label{Maxwell3.1}
V\Delta_{2}\phi-V'\Delta_{1}\phi=0
\end{equation}
\noindent Making explicit use of (\ref{aid}) and the definitions (\ref{def1}) and (\ref{def2}) we have the identities
\begin{equation}
\Delta_{1}V=h^{ij}V_{,_{i}}V_{,_{j}}=h^{ij}V'^2\phi_{,_{i}}\phi_{,_{j}}=V'^2\Delta_{1}\phi
\end{equation}
\begin{align*}
\Delta_{2}V &= h^{ij}V_{\vert\vert_{ij}}=h^{ij}(V'\phi_{\vert\vert_{ij}}+V''\phi_{,i}\phi_{,j})\\
& = V'\Delta_{2}\phi+V''\Delta_{1}\phi
\end{align*}
The above identities are now used to put equation (\ref{Einstein3.1}) in the form
\begin{equation} \label{Einstein3.2}
VV'\Delta_{2}\phi+(VV''-4\pi)\Delta_{1}\phi=0
\end{equation}
\noindent This form is in particular useful as multiplying (\ref{Maxwell3.1}) by $(-V')$  and adding the result
to (\ref{Einstein3.2}) we arrive at
\begin{equation} \label{aid2}
\Delta_{1}\phi(VV''+V'^2-4\pi)=0
\end{equation}
which is equivalent to
\begin{equation} \label{aid3}
VV''+V'^2-4\pi=0
\end{equation}
assuming $\Delta_{1}\phi\neq 0$. 
Integrated once we obtain
\begin{equation} \label{aid4}
VV'-4\pi\phi-\beta =\frac{1}{2}(V^2)'-4\pi\phi-\beta=
0
\end{equation}
\noindent where $\beta$ is an arbitrary integration constant. A second integration yields the desired relation between $V$ and $\phi$, namely \begin{equation} \label{important2}
V^2=A+B\phi+4\pi\phi^2
\end{equation}
where $A$ y $B$ are arbitrary constants. Note that we still have not used the Einstein equation (\ref{Einstein2}), and we will not in the following. Instead we assume that the electrovac universe given by the metric (\ref{electrovac}) takes a particular form of the conformastat type (\ref{conformastat}). This is to say we assume
$h_{ij}+V^{-2}\delta_{ij}$ or
\begin{equation} \label{conformastat2}
ds^2=-V^2dt^2+V^{-2}dx^{i}dx^{j}\delta_{ij}
\end{equation}
such that $V=U^{-1}$. Recalling that $T_{\mu}^{\mu}=0$ implies $R=0$, equation (\ref{important}) reduces then to the Laplace
equation
\begin{equation} \label{important3}
\Delta U\equiv U_{,_{kk}}=0
\end{equation}
\noindent
In vacuum we found that  $\sqrt{U}$ must be a harmonic function (see equation (\ref{Laplace1})). Now with (\ref{important3}) we find that $U$ satisfies Laplace equation. Therefore we have a simple way to specify an electrovac solution: take any \emph{harmonic} solution $U$ and define $V$ by $UV=1$, then use (\ref{important2}) to solve for the potential. In order to recover the flat spacetime far from the source, we must choose U such that $U^2\to 1$ at infinity. As in the vacuum case, the isotropic coordinates play a special role here and it is illuminating to dwell upon their role in the Reissner-Nordstr\"om case (which is the spherically symmetric sub-case of the more general one studied above). This metric in Schwarzschild coordinates is given by
\begin{eqnarray}
ds^2 &= -\left(1-\frac{2m}{r}+\frac{Q^2}{r^2}\right)dt^2+\left(1-\frac{2m}{r}+\frac{Q^2}{r^2}\right)^{-1}dr^2\nonumber\\
& +r^2d\Omega^2
\end{eqnarray}
The transformation to isotropic coordinate $\rho$ involves 
\begin{equation}
r=\rho\left(1+\frac{m}{\rho}+\frac{m^2-Q^2}{4\rho^2}\right)
\end{equation}     
and we obtain
\begin{eqnarray} \label{RN}
ds^2 = -\left[\frac{m^2-4\rho^2-Q^2}{(m+2\rho)^2-Q^2}\right]^2dt^2\nonumber\\
 +\left(1+\frac{m}{\rho}+\frac{m^2-Q^2}{4\rho^2}\right)^2\left(d\rho^2+\rho^2d\Omega^2\right)
\end{eqnarray}
Comparing (\ref{RN}) with (\ref{conformastat}) we can conclude that the function $U$ is
\begin{equation}
U=1+\frac{m}{\rho}+\frac{m^2-Q^2}{4\rho^2}
\end{equation}
However, a straightforward calculation tells us that
\begin{equation}
\Delta U=\frac{1}{2\rho^4}\left(m^2-Q^2\right)
\end{equation} 
Hence only if $Q=m$ holds, $U$ satisfies the Laplace equation. For this extreme case we also have
\begin{equation}
ds^2=-\left(\frac{1}{1+\frac{m}{\rho}}\right)^2dt^2+\left(1+\frac{m}{\rho}\right)^2[d\rho^2+\rho^2d\Omega^2]
\end{equation}
and therefore $V=U^{-1}$, which shows the consistency of the model.
\section{Electrovac universe with $\Lambda$}
Having reviewed the electrovac universe with vanishing cosmological constant, let us discuss a relative fast derivation of the corresponding situation where $\Lambda \neq 0$. Right from the beginning we can specialize to the conformastat case i.e.
\begin{equation}
ds^2=-f^2dt^2+f^{-2}dx^{i}dx^{j}\delta_{ij}
\end{equation}
\noindent The Einstein equation reads now 
\begin{equation}
R_{\mu\nu}-\frac{1}{2}R\text{g}_{\mu\nu}+\Lambda\text{g}_{\mu\nu}=-8\pi T_{\mu\nu}
\end{equation}
\noindent From the traceless condition of the electromagnetic tensor, we have
\begin{equation} \label{aid5}
R=4\Lambda
\end{equation} 
This can be used to re-write the Einstein equation in a form which is more suitable fur our purposes
\begin{equation}
R_{\mu\nu}-\Lambda\text{g}_{\mu\nu}=-8\pi T_{\mu\nu}
\end{equation}
This makes it evident which modification the cosmological constant $\Lambda$ introduces as compared to the results from the last section. Equation (\ref{Einstein2}) becomes
\begin{equation}
R_{ij}-\Lambda h_{ij}=-8\pi f^{-2}\left(\frac{1}{2}h_{ij}\Delta_{1}\phi-\phi_{,i}\phi_{,j}\right)
\end{equation}
whereas (\ref{Einstein3}) is simply
\begin{equation}
R_{00}+\Lambda f^2=4\pi\Delta_{1}\phi = f\Delta_{2}f
\end{equation}
Reproducing the steps from section 3, we obtain the analogy to (\ref{aid2})
\begin{equation}
f\Delta_{2}f-\Lambda f^2-4\pi\Delta_{1}\phi=0
\end{equation}
which integrated twice with respect to $\phi$ gives
\begin{equation} \label{important2x}
f^2=A+B\phi+4\pi\phi^2+\Lambda\left[\frac{1}{(\ln\phi)_{,_{i}}}\right]^2
\end{equation}
As compared to the algebraic equation (\ref{important2}), the above equation (which reduces to (\ref{important2}) in the case of $\Lambda=0$) is a non-linear partial differential equation. Finally, the combination of (\ref{important}) with (\ref{aid5}) results in
\begin{equation} \label{important3x}
\Delta U=2\Lambda U^3
\end{equation}
which is the the generalization of (\ref{important3}). The linear Laplace equation becomes now in a non-linear partial differential equation for $U$. To summarize, if we know $U$ we can use (\ref{important2x}) to infer the electric potential $\phi$ by the relation $fU=1$. We note that the level of mathematical complication introduced by $\Lambda$ is quite formidable. Note that the right side of (\ref{important3x}) shows a coupling between the cosmological constant and the function $U$ which is related to the potential $\phi$. This informs us that the electromagnetic phenomena (in our case the electric potential) will be affected by the cosmological constant. In the following section, we will offer some solutions to this equation. 
\section{Solutions}
Before we come to the non-pertubative solution we mention that an iterative one can be found by using
the standard technique. In case $\Lambda$ is small, we can attempt a pertubative solution by the ansatz
\begin{equation}
U=U_0 + \Lambda^1 U_1 + \Lambda^2 U_2 +...
\end{equation}
Back into equation (\ref{important3x}) this ansatz gives first a Laplace equation followed
by a series of Poisson equations:
\begin{eqnarray}
\Delta U_0&=&0\nonumber\\\Delta U_1&=& 2U_0^3\nonumber\\\Delta U_2&=& 6U_0^2 U_1 \nonumber\\&...&
\end{eqnarray}
Next we offer special cases of non-pertubative solutions. Let us first concentrate on the one dimension case.
In one dimension equation (\ref{important3x}) becomes an autonomous second order differential equation, namely
\begin{equation}
\frac{d^2 U}{dx^2}=2\Lambda U^3.
\end{equation}
By means of the substitution $u(U)=dU/dx$ the above equation can be reduced to the first order ODE
\begin{equation}
\frac{du^2}{dU}=4\Lambda U^3
\end{equation}
that can be integrated yielding 
\begin{equation}
u^2=\Lambda U^4+c_1
\end{equation}
where $c_1$ is an integration constant. The last step consists in integrating the ODE
\begin{equation}
\frac{dU}{dx}=\pm\sqrt{\Lambda U^4+c_1}
\end{equation}
and we obtain
\begin{equation}
c_2\pm x=\int\frac{dU}{\sqrt{\Lambda U^4+c_1}}.
\end{equation}
The integration constants $c_1$ and $c_2$ should be fixed so 
that the metric becomes de Sitter in the limit $x\to\infty$. However, the integral can be solved in terms of the elliptic function $F$ as 
\begin{equation}
c_2\pm x=\frac{1}{\sqrt{i\sqrt{\Lambda c_1}}}F\left(\sqrt{i\sqrt{\Lambda/c_1}},i\right).
\end{equation}
Independently of the sign of $c_1$ the above solution will be complex. Hence, 
the requirement that $U$ is a real function of the spatial variable $x$ will imply that $c_1=0$. 
In this case the solution is
\begin{equation}
U(x)=-\frac{1}{\sqrt{\Lambda}\left(c_2\pm x\right)}.
\end{equation}
Now, let us consider the more complicated situation where $U$ depends on two spatial variables $x$ and $y$. In this case the equation to solve is
\begin{equation}
\frac{\partial^2 U}{\partial x^2}+\frac{\partial^2 U}{\partial y^2}=2\Lambda U^3
\end{equation}
with $U=U(x,y)$. The above equation is a special case of the following more general stationary heat equation with nonlinear source, namely
\begin{equation}\label{app1}
\frac{\partial^2 U}{\partial x^2}+\frac{\partial^2 U}{\partial y^2}=f(U)\, ; \,\quad f(U)=2\Lambda U^3
\end{equation}
As in \cite{Pol} let us suppose that $U=U(x,y)$ is a solution of our equation. Then, the functions
\begin{eqnarray}
U_1=U(\pm x+C_1,\pm y+C_2)\nonumber\\U_2=U(x\cos{\beta}-y\sin{\beta},x\sin{\beta}+y\cos{\beta})
\end{eqnarray}
where $C_1$, $C_2$ and $\beta$ are arbitrary constants, are also solutions of the original equation. Implicit solutions can be found in the form
\begin{eqnarray}
\int\left[C+\frac{2}{A^2+B^2}F(U)\right]^{-1/2}=Ax+By+D\nonumber\\F(U)=\int f(U)dU
\end{eqnarray}
where $A$, $B$, $C$ and $D$ are arbitrary constants. Notice that for $f(U)=2\Lambda U^3$ the above integral gives rise
 to a complex elliptic function and again the requirement that $U$ has to be a real function fixes $C=0$ and we obtain
\begin{equation}
U(x,y)=-\sqrt{\frac{A^2+B^2}{\Lambda}}\frac{1}{Ax+By+D}.
\end{equation}
If we assume a solution with central symmetry about the point $(-C_1,-C_2)$ with $U=U(\xi)$ where
\begin{equation}
\xi=\sqrt{(x+C_1)^2+(y+C_2)^2}
\end{equation}
and $C_1$, $C_2$ are arbitrary constants, then the function $U(\xi)$ is determined by the second order non-linear differential equation
\begin{equation}
\frac{d^2 U}{d\xi^2}+\frac{1}{\xi}\frac{dU}{d\xi}=f(U).
\end{equation}
Since it is a quasi-linear equation it can be reduced to its normal form
\begin{equation}
\frac{d^2 U}{d\omega^2}=2\Lambda e^{2\omega}U^3
\end{equation}
by means of the transformation $\omega=\ln{\xi}$. If we set $2\omega=\widetilde{x}$ the previous equation becomes
\begin{equation}\label{app2}
\frac{d^2 U}{d\widetilde{x}^2}=\frac{\Lambda}{2}e^{\widetilde{x}}U^3
\end{equation}
which is a particular case of the equation
\begin{equation}
\frac{d^2 y}{dx^2}=Ae^{x}y^m\left(\frac{d y}{dx}\right)^\ell
\end{equation}
given in \cite{Pol1}. Since in our present case $\ell\neq 1-m$ we have a particular solution
\begin{equation}
U(\omega)=\frac{1}{\sqrt{2\Lambda}}e^{-\omega}.
\end{equation}
On the other side $m\neq 0$ and $\ell\neq 1$ and we can reduce equation (\ref{app2}) with the help of the transformation
\begin{equation}
t=\frac{dU}{d\widetilde{x}},\quad w=e^{\widetilde{x}}
\end{equation}
to a generalized Emden-Fowler equation with respect to $w=w(t)$, namely
\begin{equation}
\frac{d^2 w}{dt^2}=-3\left(\frac{\Lambda}{2}\right)^{1/3}tw^{-1}\left(\frac{dw}{dt}\right)^{7/3}.
\end{equation}
Unfortunately, the above equation does not match with those listed in \cite{Pol1}. Moreover, equation (\ref{app1}) can be seen as a particular case of 
\begin{equation}
\frac{\partial^2 U}{\partial x^2}+\frac{\partial^2 U}{\partial y^2}=aU+bU^n.
\end{equation}
For $a=0$ there is a self-similar solution of the form \cite{Pol}
\begin{equation}
U(x,y)=x^{2/(1-n)}F\left(z\right),\quad z=\frac{y}{x}.
\end{equation}
In our case for $b=2\Lambda$ and $n=3$ we shall have 
\begin{equation}
U(x,y)=x^{-1}F\left(z\right)
\end{equation}
where $F(z)$ is a solution of the second order nonlinear ODE
\begin{equation}
(1+z^2)\frac{d^2 F}{dz^2}+4z\frac{dF}{dz}+2F=2\Lambda F^3.
\end{equation}
The solution of the above equation can be expressed in terms of the Jacobi amplitude function $J_{SN}$ as follows
\begin{eqnarray}
F(z)=\frac{A_2}{\sqrt{(1-\Lambda+A_2^2\Lambda)(1+z^2)}}\times\nonumber\\
J_{SN}\left(\frac{\sqrt{1-\Lambda}\arctan(z)+A_1}{\sqrt{1-\Lambda+A_2^2\Lambda}},A_2\frac{\sqrt{\Lambda(1-\Lambda)}}{\Lambda-1}\right)
\end{eqnarray}
Finally, equation (\ref{important3x}) can be seen as a special case of the more general equation
\begin{equation}
\frac{\partial^2 U}{\partial x^2}+\frac{\partial^2 U}{\partial y^2}=aU^{n}+bU^{2n-1}
\end{equation}
with $a=2\Lambda$, $n=3$ and $b=0$. For this choice the solutions of the above equation are \cite{Pol}
\begin{equation}
U(x,y)=\left[\frac{\Lambda}{2}(x\sin{\alpha_1}+y\cos{\alpha_1}+\alpha_2)\right]^{-1/2}
\end{equation}
and
\begin{equation}\label{final}
U(x,y)=\frac{1}{\sqrt{2\Lambda\left[(x+\alpha_1)^2+(y+\alpha_2)^2\right]}}
\end{equation}
where $\alpha_1$ and $\alpha_2$ are arbitrary constants. In contrast to the vanishing $\Lambda$ case, here we must recover the de Sitter spacetime at infinity. Note that (\ref{final}) shows in explicit form the idea discussed previously about the coupling between electromagnetism and cosmology in this theory, i.e., the cosmological constant affects the local electromagnetic phenomena, considering that $U$ (which now is a function of $\Lambda$) will determine the potential $\phi$. Notice that except for the case of a self-similar solution the above results can be easily generalized to the case when $U$ depends on all three spatial variables.\\
\section{Conclusions}
We have discussed in detail the derivation of the differential equations governing the electrovac universes model with a cosmological constant $\Lambda$. These equations in the case when $\Lambda=0$ consist of an algebraic equation of second order and a Laplace equation. This model provides a simple way to construct electrovac solutions: take a harmonic function U, and then obtain the electric potential $\phi$ by using (\ref{important2}). The mathematical complexity of the model increases when $\Lambda$ is introduced, considering that the fundamental equations both become non-linear differential equations. Far away, the theory becomes more interesting when $\Lambda$ enters in the scenario, because it shows from the generalized non-linear Laplace equation (\ref{important3x}), the coupling between the cosmological constant and the function $U$ which determines the potential $\phi$. Is worth to remark here that in this model the cosmological constant affects the local electromagnetic phenomena. We showed that (\ref{important3x}) can be solved using different techniques discussed in the last section. We offer several particular solutions of (\ref{important3x}) in one and two dimensions and comment on the possible generalization to three dimensions. These solutions shows the central idea of the electrovac universes with $\Lambda$: the electric potential $\phi$ will be affected by the cosmological constant, provided that U is now a function of $\Lambda$. 
\section{Acknowledgments}
We would like to thank Dr. M. Nowakowski at Universidad de los Andes (Colombia), for inspiration for this work and useful discussions. N.C.P. wants to thank to the Physics Department at the Universidad de los Andes for its support while the bulk of this work was completed. N.C.P also wants to thank to Prof. Frank Avignone III (USC) for his support during the time this work was culminated. 

\end{document}